\documentclass[11pt]{article}

\usepackage{geometry}
\usepackage{amsthm}
\usepackage{amsfonts,colortbl}
\usepackage{amsmath,bm,hyperref,multirow}
\usepackage{graphicx}
\usepackage{caption}
\usepackage{subcaption}
\usepackage{xcolor}

 \RequirePackage{threeparttable}
 
 \newcommand{\headrow}{\rowcolor{gray}}
 \newcommand{\thead}[1]{\multicolumn{1}{l}{\bfseries #1\rule[-1.2ex]{0pt}{2em}}}

\newtheorem{theorem}{Theorem}
\newtheorem*{theorem*}{Theorem}

\newtheorem{remark}[theorem]{Remark}

\title{Consecutive \textit{k}-out-of-\textit{n}:\textit{F} Systems \\ Have Unbounded Roots}

\author{Marilena Jianu$^1$ \and Leonard D\u au\c s$^1$ \and Vlad-Florin Dr\u agoi$^2$ \and Valeriu Beiu$^2$} 

\date{\scriptsize $^1$ Department of Mathematics and Computer Science, Technical University of Civil Engineering Bucharest \newline Bucharest, 020396, Romania \\
$^2$ Department of Mathematics and Computer Science, Faculty of Exact Sciences, ``Aurel Vlaicu'' University of Arad \newline Arad, 310032, Romania}

\begin{document}

\maketitle

\begin{abstract}
This article is studying  the roots of the reliability polynomials of linear consecutive-\textit{k}-out-of-\textit{n}:\textit{F} systems. We are able to prove that these roots are unbounded in the complex plane, for any fixed $k\ge2$.  In the particular case $k=2$, we show that the reliability polynomials have only real roots and highlight the closure of these roots by establishing their explicit formulas.  We also point out that in this case, for  any fixed \textit{n} the nonzero roots of the reliability polynomial are distinct numbers. 

\medskip
% Please include a minimum of six keywords
\end{abstract}

\section{Introduction}

\textbf{\textit{Computer reliability}} was properly established around the mid '50s when John von Neumann \cite{vonNeumann} on one side, and Edward F. Moore and Claude E. Shannon \cite{MS} on the other side, set up the two main lines of thought for \textit{``building reliable systems from unreliable components.''} One line of research flourished from \textbf{\textit{networks of gates}} \cite{vonNeumann} (the components being the well-known logical gates), while the other one from \textbf{\textit{networks of devices}} \cite{MS} (the components being relays --- hence equivalent to present day transistors). Moore and Shannon approach \cite{MS} founded a probabilistic graph model where nodes (devices) are perfectly reliable, while the undirected edges (connections/wires) fail independently with a probability $q=1-p$. The fundamental problem was (and still is) that of finding the probability that two (or more) nodes are connected, the solution being represented by the \textbf{\textit{reliability polynomial}} $Rel(N,p)$ of a network $N$. 
The most common particular cases are represented by the \textit{all-terminal reliability} polynomial (which gives the probability that any two nodes are connected), and the \textit{two-terminal reliability} polynomial (which gives the probability that two fixed, particular nodes $s$ and $t$ called \textit{terminals}, are connected, also known as $s$-$t$ connectivity).

Alternatively, one could consider that the nodes are the ones failing with some probability $q=1-p$, with edges being perfectly reliable. This case is known as \textit{node reliability}, which can be either \textit{residual node connectedness} \cite{Co93} (the probability that at least one node is operational and all the operational nodes form a connected subgraph), or of the \textit{two-terminal} type \cite{BGMRNet} (the probability that two terminals $s$ and $t$ are connected, supposing that the nodes $s$ and $t$ are always operational).

The roots of \textit{all-terminal} reliability polynomials where analyzed for the first time by Jason I. Brown and Charles J. Colbourn  \cite{BC}, who conjectured (in 1992) that all these roots lie inside the closed unit disk centered at $z=1$. The conjecture was shown to be false twelve years later by very shy margins, but the original question posed by Brown and Colbourn \cite{BC} \textit{“Can we find a bounded region that is guaranteed to contain all the roots of the [all-terminal] reliability polynomials?”} is still open now (see  \cite{DG}). Since 2006 the roots of \textit{two-terminal} reliability polynomials  have also started to be scrutinized by Christian Tanguy, in a series of papers tackling particular networks \cite{2006a_Tanguy, 2006b_Tanguy,  2007b_Tanguy}. Recently,  Brown and DeGagn\' e \cite{BrownDG}  proved that the closure of two-terminal roots contains the closed unit disks centered at $z=0$ and $z=1$, but the question if the \textit{two-terminal} roots are bounded also remains an open problem (see \cite{DG, BrownDG}). 

The roots of \textit{node reliability polynomials} received far less attention. Still, Brown and Mol \cite{BMNet} proved that the roots of  \textit{residual node connectedness reliability} are unbounded and are dense in the entire complex plane, which is in stark contrast with the roots of \textit{all-terminal} and \textit{two-terminal} reliability polynomials (which, expectedly, are bounded).

As far as we know, the present paper is the first one studying the roots of \textit{two-terminal node reliability} polynomials. The particular networks we are going to consider here are known as linear consecutive $k$-out-of-$n$:$F$ systems. The linear consecutive $k$-out-of-$n$:$F$ system was introduced by Kontoleon in 1980 \cite{K80} under the name of $r$-successive-out-of-$n$:$F$, being rebranded consecutive-$k$-out-of-$n$:$F$ one year later \cite{Chiang}. It is defined as a system formed by $n$ components placed in a row (i.e., sequentially), which fails iff at least $k$ consecutive components fail. Linear consecutive-$k$-out-of-$n$:$F$ systems are of interest as they can achieve high reliability \cite{BHB} at reasonably low costs~\cite{BDB}, and have been used for street lights and oil/gas pipelines, as well as by biology \cite{DB}. 

\begin{figure}[h!]
\centering
\includegraphics[width=400pt]{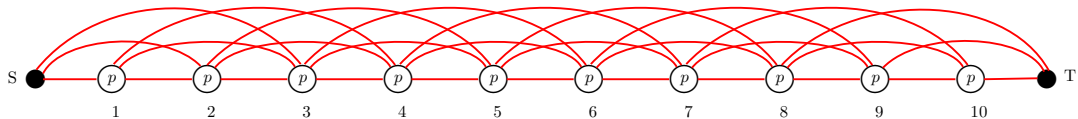}
\caption{Linear consecutive $3$-out-of-$10$:$F$ system.}
\label{Fig1}
\end{figure}

A linear consecutive-$k$-out-of-$n$:$F$ system can be represented as an undirected graph $G=(V,E)$, with $n+2$ vertices, $V=\{S=0,1,2,\ldots,n, n+1=T\}$, and having the set of  edges $E=\{i\,j,\;0\le i<j\le n+1,\;j-i\le k  \}$ (see Fig. \ref{Fig1}). The nodes $1,2,\ldots, n$ are identical and statistically independent, failing with probability $q=1-p$, while the terminals $S=0$, $T=n+1$, as well as the edges, are always operational. The reliability of the system is the probability that the system is working, being expressed as a polynomial in $p$, denoted by $Rel(k,n;p)$.

\vspace{3mm}

For any fixed $k\ge 2$, the reliability polynomials of a linear consecutive $k$-out-of-$n$ system can be computed by a recurrence relation \cite{Chang}, for all $n\ge k$. Using this recurrence relation and Beraha-Kahane-Weiss theorem, we prove, in Section 2, that the set of the roots of the polynomials $Rel(k,n;p)$ is unbounded, for any $k\ge 2$. 
% Formally, our result states as follows.

% \begin{theorem*}For any fixed  $k\ge2$, the set of complex roots of all the polynomials $Rel(k,n;p)$, $n\ge k$, is unbounded.
% \end{theorem*}
In Section 3 we study the particular case of linear consecutive $2$-out-of-$n$:$F$ systems and prove that all the nonzero roots are real, distinct numbers, and that their closure is $(-\infty, 0]\cup \left[\frac{4}{3},\infty\right)$.

\section{Consecutive-\textit{k}-out-of-\textit{n}:\textit{F} Systems Have Unbounded Roots}

For any fixed $k\ge2$, the reliability polynomials $R_n(p):=Rel(k,n;p)$ of linear consecutive $k$-out-of-$n:F$ systems satisfy the following recurrence relation \cite{Chang}:
\begin{equation}\label{EQ01}
    R_{n}(p)=pR_{n-1}(p)+pqR_{n-2}(p)+\ldots+pq^{k-1}R_{n-k}(p),
\end{equation}
for any $n\ge k$. Since the system fails only if at least $k$ consecutive components fail, the first $k$ polynomials are equal to 1:
\begin{equation}\label{EQ02}
R_n(p)=1, \;\; n=0,1,\ldots,k-1.
\end{equation}

For any $n\ge 0$, the polynomial $R_n(p)$ can be written
\begin{equation}\label{EQ03}
R_n(p)=p^n P_n(z),
\end{equation}
where 
$$
z=\frac{q}{p}=\frac{1}{p}-1.
$$
From \eqref{EQ01} we find that the polynomials $P_n(z)$ satisfy the following recurrence relation, for any $n\ge k$:
\begin{equation}\label{EQ05}
    P_{n}(z)=P_{n-1}(z)+zP_{n-2}(z)+\ldots+z^{k-1}P_{n-k}(z),
\end{equation}
and from \eqref{EQ02} we obtain the initial conditions:
\begin{equation}\label{EQ02a}
P_n(z)=(1+z)^n, \;\; n=0,1,\ldots,k-1.
\end{equation}

Consider the general case when $\{P_n\}$ is a sequence of polynomials defined by a  recurrence relation of order $k$: 
\begin{equation}\label{eq6}
P_{n}(z)=\sum_{j=1}^k f_j(z)P_{n-j}(z),\quad n\ge k,
\end{equation}
where $f_j(z)$ and $P_{j-1}(z)$, $j=1,2,\ldots,k$ are given polynomials.

We say that $x\in\mathbb C$ is a limit of zeros of $\{P_n\}$ if there exists a sequence of complex numbers $\{z_n\}$ such that $P_n(z_n)=0$ and 
$\displaystyle \lim_{n\to\infty}z_n=x$.

Let $\lambda_j(z)$, $j=1,2,\ldots,k$ be the roots of the characteristic equation 
\begin{equation}\label{eq7}
\lambda^k-\sum_{j=1}^k f_j(z)\lambda^{k-j}=0.
\end{equation}
Then, for any $z\in\mathbb C$ such that $\lambda_j(z)$ are distinct, we have:
\begin{equation}\label{eq8}
P_n(z)=\sum_{j=1}^k \alpha_j(z)\lambda_j(z)^n,
\end{equation}
where $\alpha_j(z)$, $j=1,\ldots,k$ are found from the linear system of equations obtained by  writing \eqref{eq8} for $n=0,\ldots,k-1$.

\begin{theorem}\label{T1} (Beraha-Kahane-Weiss theorem \cite{BKW1}). 
Suppose that $\{P_n\}$ is a sequence of polynomials defined by a relation of the form \eqref{eq6} such that $\{P_n\}$ satisfies no recursion of order less than $k$, and there is no constant $\omega$, with $|\omega|=1$, such that $\lambda_i=\omega\lambda_j$ for some $i\ne j$. 

Then $x$ is a limit of zeros of $\{P_n\}$ if and only if the roots of the characteristic equation \eqref{eq7} can be ordered such that one of the following holds:

(i) $\left|\lambda_1(x)\right|>\left|\lambda_j(x)\right| $ for every $j=2,\ldots,k$, and $\alpha_1(x)=0$,

(ii) $\left|\lambda_1(x)\right|=\left|\lambda_2(x)\right|=\ldots=\left|\lambda_l(x)\right|>\left|\lambda_j(x)\right| $,  $j=l+1,\ldots,k$, for some $l\ge 2$.
\end{theorem}

The sequence of polynomials $P_n(z)$ defined by the recurrence \eqref{EQ05} of order $k$, with the initial conditions \eqref{EQ02}, satisfies the conditions from Theorem \ref{T1}. In our case, the characteristic equation \eqref{eq7} is written
\begin{equation}\label{EQ07}
\lambda^k-\sum_{j=1}^k z^{j-1}\lambda^{k-j}=0.
\end{equation}

For $z=-1$, the characteristic equation \eqref{EQ07} becomes
\begin{equation}\label{EQ08}
\lambda^k-\lambda^{k-1}+\ldots +(-1)^{k-1}\lambda+(-1)^{k}=0
\end{equation}
and its roots are the roots $\neq-1$ of the equation
$$\lambda^{k+1}=(-1)^{k+1}.$$
Hence the roots of the characteristic equation \eqref{EQ08} are all distinct complex numbers, of modulus 1:  
$$\left|\lambda_1(-1)\right|=\ldots =\left|\lambda_k(-1)\right|=1.$$
By Theorem \ref{T1} case (ii), there exists a sequence of complex numbers $\{z_n\}$ such that $P_n(z_n)=0$ and 
$\displaystyle \lim_{n\to\infty}z_n=-1$.

From the recurrence relation \eqref{EQ05} one can easily deduce that
\begin{equation}\label{EQ05a}
    P_{n}(z)=(1+z)P_{n-2}(z)+  (z+z^2)P_{n-3}(z)+  \ldots+ (z^{k-2}+z^{k-1})P_{n-k}(z)+ z^{k-1}P_{n-k-1}(z),
\end{equation}
hence
$$P_n(-1)=(-1)^{k-1}P_{n-k-1}(-1),$$
for every $n\ge k+1$. Since $P_0=1$, we find that all the polynomials of the subsequence $\left\{P_{m(k+1)}\right\}_{m\ge1}$ do not have the root $z=-1$. [The same is true for the polynomials of the subsequence $\left\{P_{m(k+1)-1}\right\}_{m\ge1}$, while all the others have the root $-1$.] Therefore, the subsequence $\{z_{m(k+1)}\}_{m\ge1}$ of $\{z_n\}$ tends to  $-1$ as $m\to\infty$ and $z_{m(k+1)}\neq -1$. Since $z_{n}= \frac{1-p_{n}}{p_{n}}$, it follows that 
$$p_{m(k+1)}=\frac{1}{z_{m(k+1)}+1},\; m=1,2,\ldots$$
is a sequence of roots of the reliability polynomial $Rel(k,n;p)$,  with $\displaystyle \lim_{m\to\infty}\left|p_{m(k+1)}\right| =\infty$ and the following theorem is proved.

\begin{theorem}\label{T2}
For any fixed  $k\ge2$, the set of complex roots of all the polynomials $Rel(k,n;p)$, $n\ge k$, is unbounded.
\end{theorem}

\section{Consecutive-2-out-of-\textit{n}:\textit{F} Systems Have Real, Distinct Roots}

The two-terminal reliability polynomial of a linear consecutive-$2$-out-of-$n$:$F$ system is \cite{Chiang}
\begin{equation}\label{eq1}
Rel(2,n;p)=\sum_{j=0}^{{\left\lfloor \frac{n+1}{2}\right\rfloor}}  \binom{n-j+1}{j}p^{n-j}(1-p)^j= 
p^n\sum_{j=0}^{{\left\lfloor \frac{n+1}{2}\right\rfloor}}  \binom{n-j+1}{j}\left(\frac{1-p}{p}\right)^j=p^nP_n(z),
\end{equation}
where 
\begin{equation}\label{eq2}
z=\frac{1-p}{p}
\end{equation}
and
\begin{equation}\label{eq3}
P_n(z)=\sum_{j=0}^{\left\lfloor \frac{n+1}{2}\right\rfloor} \binom{n-j}{j}  z^j,\quad n=0,1,\ldots 
\end{equation}
(see also \eqref{EQ03}).

Any nonzero root of $Rel(2,n;p)$ corresponds to a root $z\neq -1$ of the polynomial $P_{n}(z)$, hence it follows naturally that we should study the roots of $P_n(z)$ defined by \eqref{eq3}. 

The polynomial $P_n(z)$ satisfies the recurrence relation \eqref{EQ05} for $k=2$:
\begin{equation}\label{eq4}
P_{n}(z)=P_{n-1}(z)+z\,P_{n-2}(z),\quad n=2,3,\ldots, 
\end{equation}
with the initial conditions 
\begin{equation}\label{eq5}
P_0(z)=1,\quad P_1(z)=1+z.
\end{equation}

The characteristic equation corresponding to recurrence relation \eqref{eq4} is 
\begin{equation}\label{eq10}
\lambda^2-\lambda-z=0
\end{equation}
having the roots
\begin{equation}\label{eq11}
\lambda_{1,2}(z)=\frac{1\pm \sqrt{1+4z}}{2}.
\end{equation}
So, for any $n\ge0$ and $z\ne-\frac{1}{4}$,  $P_n(z)$ can be written as
\begin{equation}
\label{eq110}
P_n(z)=\alpha_1(z)\left(\lambda_1(z)\right)^n + \alpha_2(z) \left(\lambda_2(z)\right)^n
\end{equation}
where $\alpha_{1,2}(z)$ are determined from the initial conditions \eqref{eq5}.

Using \eqref{eq110} for $n=0,1$ and the initial conditions \eqref{eq5} we find that 
$$\alpha_1(z)=\frac{ \left(\lambda_1(z)\right)^2}{\lambda_1(z)-\lambda_2(z)}\;\text{ and } \;\alpha_2(z)=-\frac{ \left(\lambda_2(z)\right)^2}{\lambda_1(z)-\lambda_2(z)},$$
hence 
\begin{equation}\label{eq12}
P_n(z)=\frac{\left(\lambda_1(z)\right)^{n+2}- \left(\lambda_2(z)\right)^{n+2}} {\lambda_1(z) - \lambda_2(z)} =\frac{1}{\sqrt{4z+1}}\left[ \left(\frac{1+\sqrt{4z+1}}{2}\right)^{n+2}-
\left(\frac{1-\sqrt{4z+1}}{2}\right)^{n+2}
\right],
\end{equation}
for every $z\ne-\frac{1}{4}$ and $n\ge0$.
If $z= -\frac{1}{4}$, then $\lambda_1\left(-\frac{1} {4}\right)=\lambda_2\left (-\frac{1} {4}\right)= \frac{1}{2}$, so $P_n\left(-\frac{1}{4} \right)=\frac{n+2}{2^{n+1}}$.

Using the form \eqref{eq12} of the polynomial $P_n(z)$ we obtain the following formula for the reliability polynomial $Rel(2,n;p)$:
\begin{equation}\label{eq012}
Rel(2,n;p)=\frac{p^n} {2^{n+2}\sqrt{\frac{4-3p}{p}}} \left[ \left(1+\sqrt{\frac{4-3p}{p}}\right)^{n+2}-
\left(1-\sqrt{\frac{4-3p}{p}}\right)^{n+2}
\right].
\end{equation}

The roots of $P_n(z)$ are the roots $z\ne-\frac{1}{4}$ of:
\begin{equation} \label{eq120}
\left(\frac{1+\sqrt{4z+1}}{1-\sqrt{4z+1}}\right)^
{n+2}=1.
\end{equation}

If we denote by 
\begin{equation} \label{eq121}
\omega_{n,j}=\cos \frac{2j\pi}{n}+i\sin\frac{2j\pi}{n},\quad j=0,1,\ldots,n-1,
\end{equation}
the $n$-th roots of unity, the roots of equation \eqref{eq120} can be written as
\begin{equation} \label{eq122}
z_{n,j}=-\frac{\omega_{n+2,j}}{\left(1+\omega_{n+2,j}\right)^2},\quad j=1,\ldots,n,\; j\ne\tfrac{n+2}{2}.
\end{equation}
Now, using \eqref{eq121} and the identity $\cos^2x=\cos^2(\pi-x)$ we get:  
\begin{equation} \label{eq123}
z_{n,j}=-\frac{1}{4\cos^2\frac{j\pi}{n+2}},\quad j=1,\ldots,\left\lfloor \tfrac{n+1}{2}\right\rfloor.
\end{equation}

It follows that all the roots of the polynomial $P_n(z)$ are real, distinct numbers  in the interval 
$\left(-\infty,-\frac{1}{4}\right)$. The set of all the roots $\left\{z_{n,j}, j=1,\ldots,\left\lfloor \tfrac{n+1}{2}\right\rfloor,n\ge 1\right\}$ is dense in this interval.

We also remark that the polynomials $J_1(z)=1$, $J_n(z)=P_{n-2}(z)$ for $n=2,3,\ldots$  are  known as \textit{Jacobstahl polynomials}, and are intimately related to the \textit{Fibonacci polynomials} (see \cite{HB78}) defined by
\begin{equation}\label{eq20}
F_{n+2}(z)=z\,F_{n+1}(z)+F_{n}(z),\quad n=1,2,\ldots,
\end{equation}
with the initial conditions 
$F_1(z)=1,\;F_2(z)=z.$
Fibonacci and Jacobsthal polynomials have the same set of coefficients (see Table \ref{Table1}), being related as
\begin{equation}\label{eq23}
F_n(z)=z^n\,J_{n}\left(\frac{1}{z^2}\right).
\end{equation}

\begin{table}[h]
\caption{Fibonacci and Jacobsthal polynomials}
\begin{threeparttable}
\begin{tabular}{lll}
\headrow
\thead{} & \thead{Fibonacci polynomials} & \thead{Jacobsthal polynomials}\\
%\hiderowcolors
  & $F_{n+2}(z)=zF_{n+1}(z)+F_n(z)$ & $J_{n+2}(z)=J_{n+1}(z)+zJ_n(z)$\\
$n=1$ & $F_1(z)=1$ & $J_1(z)=1$\\
$n=2$ & $F_{2}(z)=z$ & $J_{2}(z)=P_0(z)=1$\\
$n=3$ & $F_{3}(z)=z^2+1$ & $J_{3}(z)=P_1(z)=1+z$\\
$n=4$ & $F_{4}(z)=z^3+2z$ & $J_{4}(z)=P_2(z)=1+2z$\\
$n=5$ & $F_{5}(z)=z^4+3z^2+1$ & $J_{5}(z)=P_3(z)=1+3z+z^2$\\
$n=6$ & $F_{6}(z)=z^5+4z^3+3z$ & $J_{6}(z)=P_4(z)=1+4z+3z^2$\\
$n=7$ & $F_{7}(z)=z^6+5z^4+6z^2+1$ & $J_{7}(z)=P_5(z)=1+5z+6z^2+z^3$\\
%$n=8$ & $F_{8}(z)=z^7+6z^5+10z^3+4z$ & $J_{8}(z)=P_6(z)=1+6z+10z^2+4z^3$\\
\hline  % Please only put a hline at the end of the table
\end{tabular}
\end{threeparttable}
\label{Table1}
\end{table}

It is known that the roots of the the Fibonacci polynomials $F_n(z)$ are given by the formula (see \cite{HB73}):
\begin{equation}\label{eq24}
w_{n,j}=2i\cos\frac{j\pi}{n},\;\; j=1,2,\ldots,n-1.
\end{equation}
Since $P_{n}(z)=J_{n+2}(z)$, $n=0,1,...$, and using \eqref{eq23} we recover once again formula \eqref{eq123} for the roots $z_{n,j}$ of the polynomial $P_{n}(z)$.
 
Now, by relying on eqs. \eqref{eq2} and \eqref{eq123}, the nonzero roots of $Rel(2,n;p)$
 can be written as:
\begin{equation}\label{eq26}
p_{n,j}=\frac{1}{1+z_{n+1,j}}=1+\frac{1}{4\cos^2\frac{j\pi}{n+2}-1},\;\; j=1,2,\ldots,\left\lfloor\frac{n+1}{2}\right\rfloor,\; j\ne\frac{n+2}{3},
\end{equation}
and so we have just proved  Theorem \ref{T3}.

\begin{theorem}\label{T3}
For any $n\ge 2$, the nonzero roots of the reliability polynomial $Rel(2,n;p)$  are real, distinct numbers.

If $\mathcal R$ denotes the set of all the roots of the polynomials $Rel(2,n;p)$, $n\ge2$, then the closure of $\mathcal R$ is: 
$$\overline{\mathcal R}= (-\infty,0]\cup\left[\frac{4}{3},\infty\right).$$
\end{theorem}

\begin{figure}[!ht]
    \centering
    \includegraphics[width=0.6\textwidth]{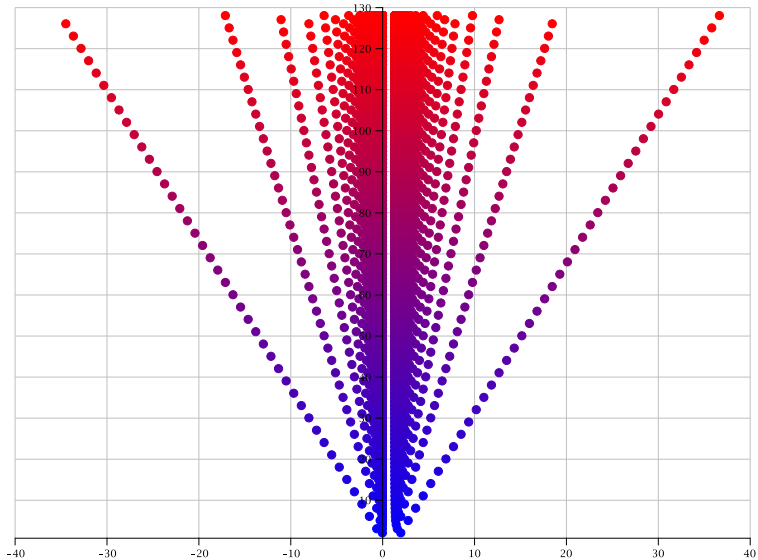}
    \caption{The roots of the reliability polynomials $Rel(2,n;p)$ for $n=2,\ldots,128$.}
    \label{fig:rootsk2}
\end{figure}

We present in Fig. \ref{fig:rootsk2} all the roots of the 127 polynomials $Rel(2,n;p)$ for $n=2,\ldots,128$ (the vertical axis being $n$).

\begin{remark}For each $n\ge 2$, the largest root of $Rel(2,n;p)$ (in absolute value) is obtained for those $j\ne\frac{n+2}{3}$ that minimize the denominator  $\left| \cos^2\frac{j\pi}{n+2}-\frac{1}{4}\right|$:
$$j=\left\{\begin{array}{cc}
   \left\lfloor\frac{n}{3}\right\rfloor+1,  & \text{if } n\not\equiv 1 \pmod 3  \\
       \left\lceil\frac{n}{3}\right\rceil+1,  & \text{if } n \equiv 1 \pmod 3
\end{array} .  \right .$$
\end{remark}

\section{Conclusions}

In this paper we have proved that the roots of linear consecutive $k$-out-of-$n$:$F$ systems are unbounded. This implies that the roots of two-terminal node reliability polynomials are unbounded.

In particular, we have studied the reliability polynomials corresponding to linear consecutive $2$-out-of-$n$:$F$ systems, showing that all their nonzero roots are distinct, real numbers, and are forming a dense set in $(-\infty,0]\cup\left[\frac{4}{3},\infty\right)$.

\begin{figure}[!ht]
     \centering
     \begin{subfigure}[b]{0.45\textwidth}
         \centering
         \includegraphics[width=\textwidth]{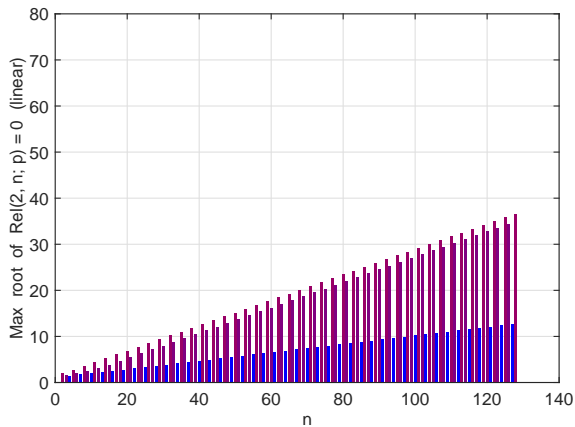}
         \caption{Linear consecutive}
         \label{fig:Consec2a}
     \end{subfigure}
     \hfill
     \begin{subfigure}[b]{0.45\textwidth}
         \centering
         \includegraphics[width=\textwidth]{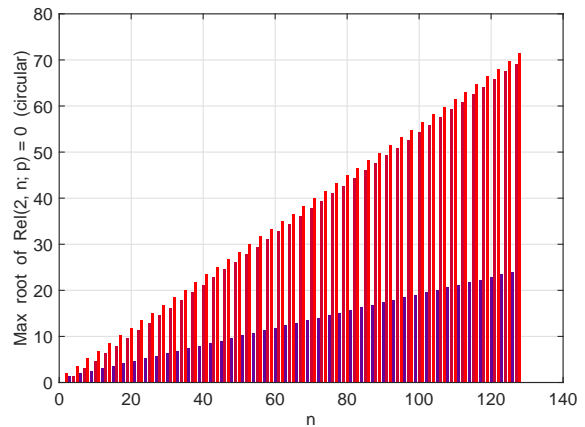}
         \caption{Circular consecutive}
         \label{fig:Consec2b}
     \end{subfigure}
        \caption{Largest roots of $Rel(2,n;p)$ for $n=2,...,128$.}
        \label{fig:Consec2}
\end{figure}

The next step we plan to take is to have a closer look at circular consecutive-$k$-out-of-$n$:$F$ systems. This choice is motivated by simulations of both linear and circular consecutive-$2$-out-of-$n$:$F$ systems, which suggest that the roots of circular consecutive-$2$-out-of-$n$:$F$ systems are growing even faster than those of linear consecutive-$k$-out-of-$n$:$F$ systems (the largest roots for $n=2,...,128$ are presented in Fig. \ref{fig:Consec2} ).

\section*{\normalsize{ACKNOWLEDGMENTS}}
This research was partly funded by a grant of the Romanian Ministry of Education and Research, CNCS-UEFISCDI, project no. PN-III-P4-ID-PCE-2020-2495, within PNCDI III (ThUNDER$^2$ = \textit{Techniques for Unconventional Nano-Designing in the Energy-Reliability Realm}).

%REFERENCES
%Note that references are placed in lexicographic order by author name. Multiple entries within a citation should appear in numerical order also, such as [2, 17, 19, 42].

\bibliographystyle{plain}
\bibliography{sample}

\end{document}